# On the Frame Error Rate of Transmission Schemes on Quasi-Static Fading Channels

Ioannis Chatzigeorgiou, Ian J. Wassell
Digital Technology Group, Computer Laboratory
University of Cambridge, United Kingdom
Email: {ic231, ijw24}@cam.ac.uk

Rolando Carrasco
School of EE & Comp. Engineering
University of Newcastle, United Kingdom
Email: r.carrasco@ncl.ac.uk

*Abstract*— It is known that the frame error rate of turbo codes on quasi-static fading channels can be accurately approximated using the convergence threshold of the corresponding iterative decoder. This paper considers quasi-static fading channels and demonstrates that non-iterative schemes can also be characterized by a similar threshold based on which their frame error rate can be readily estimated. In particular, we show that this threshold is a function of the probability of successful frame detection in additive white Gaussian noise, normalized by the squared instantaneous signal-to-noise ratio. We apply our approach to uncoded binary phase shift keying, convolutional coding and turbo coding and demonstrate that the approximated frame error rate is within 0.4 dB of the simulation results. Finally, we introduce performance evaluation plots to explore the impact of the frame size on the performance of the schemes under investigation.

## I. INTRODUCTION

The performance analysis of transmission schemes on quasi-static fading channels constitutes an important problem owning to the fact that this channel model characterizes practical settings that experience extremely slow fading conditions, such as fixed wireless access systems [1]. In this context, bounding techniques for the error rate of various transmission schemes have been proposed; such schemes include block codes [2], convolutional codes [3], [4], turbo codes [5], space-time trellis codes [6], [7] and serially concatenated codes [5], [8].

El Gamal and Hammons [9] have proposed an analytical approximation to the frame error rate (FER) of iterative schemes, such as turbo codes, on quasi-static fading channels; this tight approximation is made possible owning to a simple characterization of an iterative decoder. In this paper, we will demonstrate that non-iterative schemes, both uncoded and coded, can also be characterized in a similar manner, hence their FER performance can be estimated using the same approximation technique.

The rest of the paper is organized as follows. The quasi-static channel model as well as the standard technique to compute the FER performance of a scheme on that channel are briefly described in Section II. In Sections III-V, we derive expressions based on which a threshold value, characteristic of the system, can be determined that allows the computation of a simple yet accurate approximation to the FER. Numerical results are presented in Section VI, whilst a method to produce performance evaluation plots is discussed in VII. Finally, the main conclusions of our work are summarized in Section VIII.

## II. SYSTEM MODEL

If $\mathbf{x}$ is a frame of symbols transmitted over a quasi-static fading channel at a particular time instant and $\mathbf{y}$ is the receive frame, the input-output relationship of the channel is given by

$$\mathbf{y} = h\mathbf{x} + \mathbf{n}. \quad (1)$$

The instantaneous fading coefficient $h$ is a zero-mean, circularly symmetric complex Gaussian random variable with variance $\sigma^2 = 1$, whilst $\mathbf{n}$ is a sequence of zero-mean, mutually independent, circularly symmetric complex Gaussian random variables with variance $\mathcal{N}_0$. Note that $h$ is constant for the duration of the transmit frame but changes independently from frame to frame.

The quality of a quasi-static fading channel is characterized by its corresponding average receive signal-to-noise ratio (SNR). In particular, if $\mathcal{E}_s$ is the energy per transmit symbol and $\gamma = |h|^2 \mathcal{E}_s / \mathcal{N}_0$ is the instantaneous receive SNR, the average SNR, $\bar{\gamma}$, at the input of the receiver is given by

$$\bar{\gamma} = \mathbb{E}\left[\gamma\right] = \mathbb{E}\left[|h|^2\right](\mathcal{E}_s/\mathcal{N}_0) = \sigma^2(\mathcal{E}_s/\mathcal{N}_0) = \mathcal{E}_s/\mathcal{N}_0, \quad (2)$$

where $\mathbb{E}[.]$ denotes the expectation operator.

The average FER on a quasi-static fading channel, denoted as $P_e^Q(\bar{\gamma})$, can be computed by integrating the FER in additive white Gaussian noise (AWGN), represented by $P_e^G(\gamma)$, over the fading distribution [10]

$$P_e^Q(\bar{\gamma}) = \int_0^\infty P_e^G(\gamma) p_{\bar{\gamma}}(\gamma) d\gamma. \quad (3)$$

Now, the fading magnitude $|h|$ has a Rayleigh distribution, so that the instantaneous value of $\gamma$ is chi-squared distributed with two degrees of freedom [10], i.e.,

$$p_{\bar{\gamma}}(\gamma) = (1/\bar{\gamma}) e^{-\gamma/\bar{\gamma}}, \quad \text{for } \gamma \geq 0. \quad (4)$$

Although (3) is an exact expression for $P_e^Q(\bar{\gamma})$, its evaluation could prove difficult depending upon the transmission technique under consideration. In the following section we describe a simple approach that is often used to bound the FER performance of a communication scheme.

## III. APPROXIMATION TO THE FRAME ERROR RATE

Given an arbitrary SNR threshold $\gamma_w$, we can rewrite the expression for the average FER on a quasi-static fading channel as follows

$$P_e^Q(\bar{\gamma}) = P(\text{error}|\gamma \leq \gamma_w)P(\gamma \leq \gamma_w) \\ + P(\text{error}|\gamma > \gamma_w)P(\gamma > \gamma_w). \quad (5)$$

The SNR threshold $\gamma_w$, which we refer to as the *waterfall threshold*, is used to divide the range of SNR values into a low-SNR region and a high-SNR region. A common approach to simplify (5) is to use a trivial bound of the form

$$P(\text{error}|\gamma \leq \gamma_w) \leq 1 \quad (6)$$

for the low-SNR region and a conventional union bound for the high-SNR region. This approach has been used to bound the performance of many transmission schemes on quasi-static fading channels, including convolutional codes [4] and turbo codes [5]. Nevertheless, the value of $\gamma_w$ needs to be chosen appropriately to make the bound as tight as possible. The optimization process presented in [4], [5] and [8] produced quite tight bounds but showed that there is room for improvement.

Turbo codes have also been considered in [9] and [11]; the authors further simplified (5) by assuming that

$$P(\text{error}|\gamma \leq \gamma_w) \approx 1 \text{ and } P(\text{error}|\gamma > \gamma_w) \approx 0. \quad (7)$$

Substituting (7) into (5) gives an approximation to the FER, denoted as $\tilde{P}_e^Q(\bar{\gamma}, \gamma_w)$. In particular,

$$\begin{aligned} P_e^Q(\bar{\gamma}) &\simeq P(\gamma \leq \gamma_w) \\ &= \int_0^{\gamma_w} p_{\bar{\gamma}}(\gamma) d\gamma \\ &= 1 - e^{-\gamma_w/\bar{\gamma}} \\ &\triangleq \tilde{P}_e^Q(\bar{\gamma}, \gamma_w). \end{aligned} \quad (8)$$

It has been shown in [9] that $\tilde{P}_e^Q(\bar{\gamma}, \gamma_w)$ very accurately describes the actual FER of a turbo code using a long interleaver on a quasi-static fading channel, if the waterfall threshold $\gamma_w$ is set to be equal to the decoder convergence threshold $\gamma_{th}$. The convergence threshold of iterative schemes, such as turbo codes, can be determined using extrinsic information (EXIT) chart analysis [12].

Motivated by the work of Bouzekri and Miller [4], [5], El Gamal and Hammons [9] and Rodrigues *et al.* [11], we assume that for any transmission scheme, whose conditional FER can be reasonably described by (7), there is a value of $\gamma_w$ for which the average FER, $P_e^Q(\bar{\gamma})$, can be accurately approximated by $\tilde{P}_e^Q(\bar{\gamma}, \gamma_w)$. Based on that assumption, we derive an exact expression for the waterfall threshold in the following section.

## IV. EVALUATION OF THE WATERFALL THRESHOLD

Let $\varepsilon$ denote the absolute difference between the actual frame error probability $P_e^Q(\bar{\gamma})$ and the approximated frame error probability $\tilde{P}_e^Q(\bar{\gamma}, \gamma_w)$, i.e.,

$$\varepsilon = \left| P_e^Q(\bar{\gamma}) - \tilde{P}_e^Q(\bar{\gamma}, \gamma_w) \right|, \quad (9)$$

where $\bar{\gamma}$ can be any nonnegative real number. In the previous section we indicated that we expect the approximated FER of a transmission scheme on a quasi-static fading channel to closely represent the actual FER for a very wide range of $\bar{\gamma}$ values, provided that an appropriate value for the waterfall threshold is chosen. Consequently, if $\tilde{P}_e^Q(\bar{\gamma}, \gamma_w)$ perfectly coincides with $P_e^Q(\bar{\gamma})$, expression (9) simplifies to

$$\varepsilon = P_e^Q(\bar{\gamma}) - \tilde{P}_e^Q(\bar{\gamma}, \gamma_w) = 0. \quad (10)$$

In this section we derive an expression for the waterfall threshold $\gamma_w$ under the assumption that $\varepsilon = 0$, while in Section VI we compare our analytic approach to simulation results in order to test the validity of our assumption.

We set $\lambda = 1/\bar{\gamma}$ and express $P_e^Q(\bar{\gamma})$ and $\tilde{P}_e^Q(\bar{\gamma}, \gamma_w)$ as functions of $\lambda$, i.e., $P_e^Q(\lambda)$ and $\tilde{P}_e^Q(\lambda, \gamma_w)$ respectively. The change of variable will not have any effect on $\varepsilon$, hence

$$\varepsilon = P_e^Q(\lambda) - \tilde{P}_e^Q(\lambda, \gamma_w) = 0, \quad (11)$$

for all values of $\lambda \geq 0$. Equivalently, the area under the graph of $P_e^Q(\lambda)$ should be equal to the area under $\tilde{P}_e^Q(\lambda, \gamma_w)$, for $\lambda \in [0 \ldots \Lambda]$, where $\Lambda \to \infty$. Consequently, we can write

$$\lim_{\Lambda \to \infty} \left\{ \int_0^{\Lambda} P_e^Q(\lambda) d\lambda - \int_0^{\Lambda} \tilde{P}_e^Q(\lambda, \gamma_w) d\lambda \right\} = 0. \quad (12)$$

Using (3) and (4), we expand the first integral in (12) into

$$\begin{aligned} \int_0^{\Lambda} P_e^Q(\lambda) d\lambda &= \int_0^{\Lambda} \int_0^{\infty} P_e^G(\gamma) p_{\lambda}(\gamma) d\gamma d\lambda \\ &= \int_0^{\infty} P_e^G(\gamma) \int_0^{\Lambda} \lambda e^{-\lambda \gamma} d\lambda d\gamma. \end{aligned} \quad (13)$$

The frame error probability of a transmission scheme over an AWGN channel can also be expressed as $P_e^G(\gamma) = 1 - P_d^G(\gamma)$, where $P_d^G(\gamma)$ is the probability of successful frame detection. Consequently, we can rewrite (13) as

$$\begin{aligned} \int_0^{\Lambda} P_e^Q(\lambda) d\lambda &= \int_0^{\infty} \int_0^{\Lambda} \lambda e^{-\lambda \gamma} d\lambda d\gamma - \\ &\quad - \int_0^{\infty} P_d^G(\gamma) \int_0^{\Lambda} \lambda e^{-\lambda \gamma} d\lambda d\gamma. \end{aligned} \quad (14)$$

Taking into account that [13]

$$\int_0^{\Lambda} \lambda e^{-\lambda \gamma} d\lambda = \frac{1}{\gamma^2} - \frac{e^{-\Lambda \gamma}}{\gamma^2}(1 + \Lambda \gamma) \quad (15)$$

and

$$\int_0^{\infty} \int_0^{\Lambda} \lambda e^{-\lambda \gamma} d\lambda d\gamma = \Lambda, \quad (16)$$

the first integral in (12) assumes the form

$$\begin{aligned} \int_0^{\Lambda} P_e^Q(\lambda) d\lambda &= \Lambda - \int_0^{\infty} \frac{P_d^G(\gamma)}{\gamma^2} d\gamma + \\ &\quad + \int_0^{\infty} P_d^G(\gamma) \frac{e^{-\Lambda \gamma}}{\gamma^2}(1 + \Lambda \gamma) d\gamma. \end{aligned} \quad (17)$$

The second integral in (12) can be evaluated as follows

$$-\int_0^\Lambda \tilde{P}_e^Q(\lambda, \gamma_w)d\lambda = -\int_0^\Lambda \left(1 - e^{-\lambda\gamma_w}\right) d\lambda = -\Lambda - \frac{e^{-\Lambda\gamma_w}}{\gamma_w} + \frac{1}{\gamma_w}. \quad (18)$$

If we substitute (17) and (18) into (12), we observe that terms $\Lambda$ and $-\Lambda$ cancel each other out. Furthermore, if we take the limit as $\Lambda \to \infty$, all terms containing $e^{-\Lambda}$ are eliminated since $e^{-\Lambda} \to 0$. The remaining terms give

$$-\int_0^\infty \frac{P_d^G(\gamma)}{\gamma^2}d\gamma + \frac{1}{\gamma_w} = 0, \quad (19)$$

which is equivalent to

$$\gamma_w = \left(\int_0^\infty \frac{P_d^G(\gamma)}{\gamma^2}d\gamma\right)^{-1}. \quad (20)$$

We have thus shown that, under assumption (10), the waterfall threshold is inversely proportional to the area under a curve, which is defined by the probability of successful frame detection in AWGN normalized by the squared instantaneous SNR.

Depending on the expression for the detection probability $P_d^G(\gamma)$, a closed-form solution for $\gamma_w$ for a particular transmission technique may not exist. In that case, $\gamma_w$ can be evaluated either via numerical integration if there is an exact representation of $P_d^G(\gamma)$ or via Monte Carlo simulation if $P_d^G(\gamma)$ cannot be accurately evaluated. In the following section we revisit (20) to obtain a more practical expression for $\gamma_w$ when Monte Carlo simulation is required.

## V. Practical Computation of the Waterfall Threshold

Let $\gamma'$ be the actual SNR value for which $P_d^G(\gamma) = 0$ if $\gamma < \gamma'$ but $P_d^G(\gamma) > 0$ otherwise. Based on the definition of $\gamma'$, we can rewrite (20) as follows

$$\gamma_w = \left(\int_{\gamma'}^\infty \frac{P_d^G(\gamma)}{\gamma^2}d\gamma\right)^{-1} \quad (21)$$

or, equivalently,

$$\gamma_w = \left(\frac{1}{\gamma'} - \int_{\gamma'}^\infty \frac{P_e^G(\gamma)}{\gamma^2}d\gamma\right)^{-1}. \quad (22)$$

It becomes evident in Fig. 1 that as $\gamma$ grows, function $P_d^G(\gamma)/\gamma^2$ gradually approaches $1/\gamma^2$, which slowly converges towards zero. The advantage of (22) over (20) is that $P_e^G(\gamma)/\gamma^2$ converges to zero much faster than $P_d^G(\gamma)/\gamma^2$, hence an accurate value for $\gamma_w$ can be obtained by considering only a limited low-SNR range of integration.

Let us now consider the case when Monte Carlo simulation is used to measure the FER performance of a transmission scheme in AWGN. We assume that the SNR values $\gamma_i$, with $i = 1, 2, \ldots, N$, are equally spaced and ordered, while the FER is $P_e^G(\gamma_i) = 1$ for $i < k$ but $P_e^G(\gamma_i) < 1$ otherwise. Elaborating

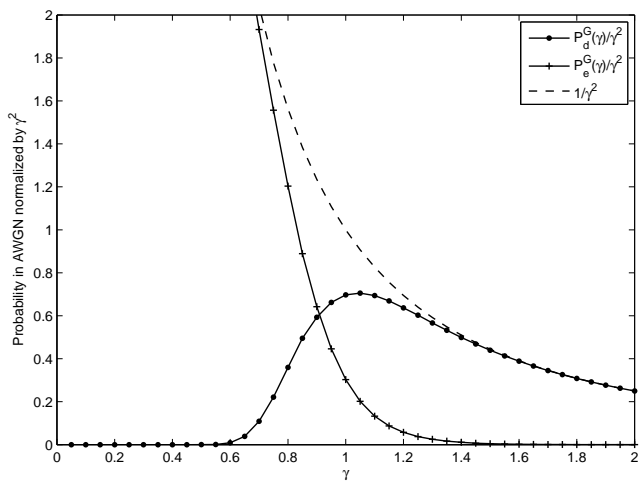

Fig. 1. Normalized probabilities in AWGN. In this example, we have considered an input frame length of 512 bits and a system using a recursive convolutional code with generator polynomials (1,17/15) in octal form.

on (22), we can obtain the following equivalent expression for discrete SNR values

$$\begin{aligned}\gamma_w &= \left(\frac{1}{\gamma_k - \left(\frac{\gamma_k - \gamma_{k-1}}{2}\right)} - \sum_{i=k}^N \frac{P_e^G(\gamma_i)}{\gamma_i^2}\right)^{-1} \\ &= \left(\frac{2}{\gamma_{k-1} + \gamma_k} - \sum_{i=k}^N \frac{P_e^G(\gamma_i)}{\gamma_i^2}\right)^{-1}.\end{aligned} \quad (23)$$

Substituting (23) into (8) gives us an approximation of the average FER of the transmission scheme on a quasi-static fading channel.

Note that the complexity of the threshold-based FER computation using (8) and either (20) or (23) is markedly less than that of the exact FER computation based on (3), as we will now demonstrate. Let us consider the case when the frame error probability in AWGN, $P_e^G(\gamma)$, is known for $N$ values of $\gamma$ and our objective is to compute the FER for a quasi-static fading channel, $P_e^Q(\bar{\gamma})$, for $M$ values of $\bar{\gamma}$. The threshold-based approach involves the evaluation of the waterfall threshold, which has a computational complexity proportional to $N$, followed by the calculation of the FER approximation, which has a computational complexity proportional to $M$. Consequently, the overall complexity of the threshold-based FER computation is of order $\mathcal{O}(N+M)$. In contrast, computation of the exact FER requires $N$ multiplications for each value of $\bar{\gamma}$, resulting in a complexity order of $\mathcal{O}(NM)$.

## VI. Numerical Results

In this section we compare analytical to simulation results for transmission over quasi-static fading channels. We consider both uncoded and coded binary phase shift keying (BPSK); coded transmission uses either a rate 1/2 recursive systematic convolutional code with octal generator polynomials (1,17/15) or a rate 1/3 turbo code with generator polynomials (1,5/7,5/7). The input frame length $L$ is either 256 bits or 1024 bits.

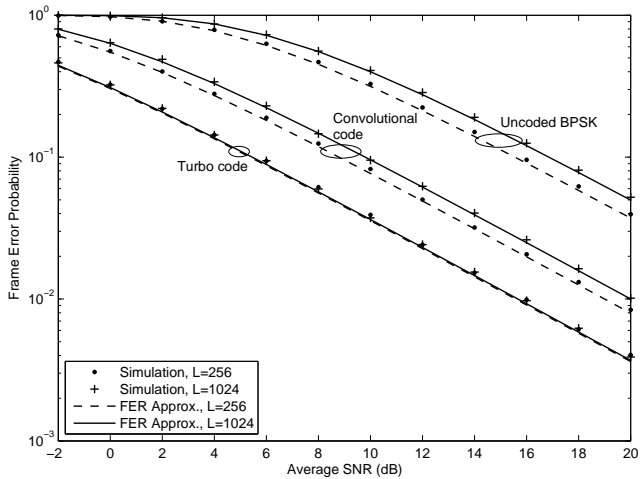
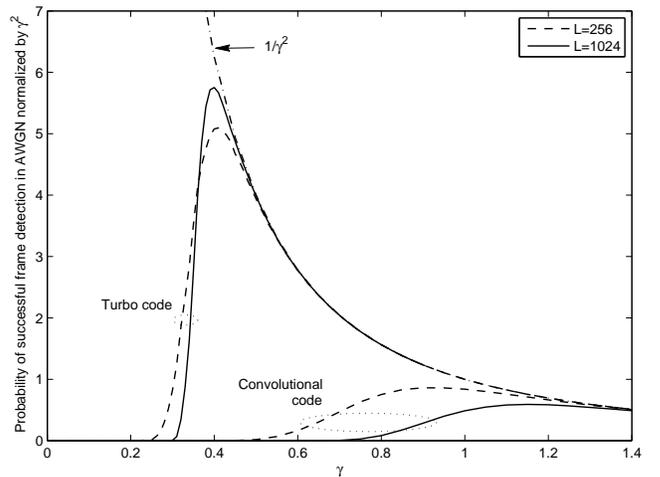

Fig. 2. Frame error rate performance of various transmission schemes on a quasi-static fading channel for input frame lengths of 256 and 1024 bits.

Fig. 3. Performance plots for the convolutional and turbo codes under consideration.

The waterfall threshold of uncoded BPSK was computed numerically using (20), where the probability of successful frame detection is captured by [14]

$$P_d^G(\gamma) = \left(1 - Q(\sqrt{2\gamma})\right)^L. \qquad (24)$$

Here, $Q(x)$ is the tail integral of a standard Gaussian density with zero mean and unit variance, defined as

$$Q(x) = \int_x^\infty \left(1/\sqrt{2\pi}\right) e^{-u^2/2} du. \qquad (25)$$

The threshold $\gamma_w$ was found to be 5.782 dB for $L=256$ and 7.083 dB for $L=1024$.

In the case of coded transmission, the probability of erroneous frame detection in AWGN was obtained running Monte Carlo simulations for a limited range of low SNR values (e.g., $\gamma \in (0,1]$ for turbo coding) and a small number of channel realizations (a few thousands at most). The waterfall threshold was then calculated using (23). In particular, when convolutional coding is employed it was found that $\gamma_w = -0.983$ dB for $L=256$, while $\gamma_w = 0.023$ dB for $L = 1024$. When turbo coding is used, however, the frame length appears to have minimal impact on the waterfall threshold; its value was found to be $\gamma_w = -4.401$ dB when $L=256$ and $\gamma_w = -4.312$ dB when $L = 1024$. Hence, we expect that the input frame length will not significantly affect the FER performance of the turbo code.

Once the waterfall threshold of a scheme has been computed, we substitute it into (8) to obtain an analytical expression for the approximated average FER for the quasi-static fading channel. The curves of the approximated FER expressions for the systems under investigation are compared to simulation results in Fig. 2. Observe that the analytic technique very closely approximates (within 0.4 dB) the simulation results in the various scenarios. Hence, both coded and uncoded, iterative and non-iterative systems can indeed be characterized by waterfall thresholds based on which tight approximations for their frame error probability can be derived. It is also interesting to note that, as expected, the FER performance of the turbo code remains unaffected by the input frame length, or equivalently the interleaver size. The same behavior has also been reported in [5], [11].

Based on the comparison between analytical and simulation results, we conclude that the technique presented in this paper accurately estimates the FER performance of BPSK transmission schemes over quasi-static fading channels. Similar results can be easily obtained using the same approach for higher order modulations.

## VII. DISCUSSION ON PERFORMANCE EVALUATION

An insight into the relative performance of two or more transmission schemes on quasi-static fading channels can also be obtained by plotting their normalized probabilities of successful frame detection in AWGN, i.e, $P_d^G(\gamma)/\gamma^2$, and comparing the areas under the corresponding curves. In particular, the larger an area is, the smaller is $\gamma_w$ and, consequently, the better the FER of the scheme under consideration is expected to be, according to (8) and (20).

The normalized probabilities $P_d^G(\gamma)/\gamma^2$ of the coded BPSK schemes that we considered in the previous section, namely the rate 1/2 convolutional code and the rate 1/3 turbo code, have been plotted in Fig. 3. We observe that as the input frame length of the convolutional code increases, the area under the normalized probability graph reduces and, as we have already seen in Fig. 2, the FER performance of the convolutionally coded scheme degrades. As anticipated, the area under the normalized probability curve of the turbo code is clearly larger than that of the convolutional code, hence the former scheme yields better FER performance on quasi-static fading channels. Most importantly however note that, in the case of the turbo code, as the length of the input frame increases, the shape of the curve changes such that its peak shifts leftward but at the same time, moves upward. For $L$ large, it is expected that the peak will reach its maximum value, which is a point on $1/\gamma^2$;

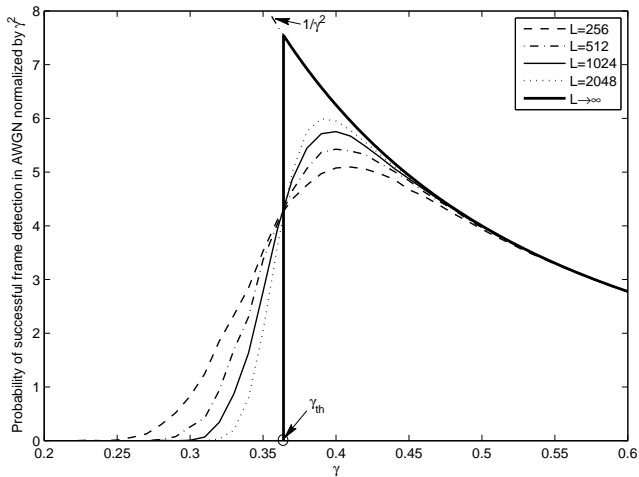

Fig. 4. Performance plots of a turbo code with generator polynomials (1,5/7,5/7), for various interleaver sizes.

notice that the curve $1/\gamma^2$ corresponds to the ideal case where $P_d^G(\gamma) = 1$ for all values of $\gamma$.

This trend is more clearly illustrated in Fig. 4, where the normalized probability of the turbo code has been plotted for various input frame lengths or, equivalently, interleaver sizes. In all cases, the exact log-MAP algorithm was used, while all probabilities were measured after 8 decoding iterations. We can infer from Fig. 4 that the probability of successful frame detection of the turbo code becomes a step function when very large interleavers are used, i.e., $L \to \infty$. In particular, $P_d^G(\gamma) = 0$ for $\gamma < \gamma_{th}$ whereas $P_d^G(\gamma) = 1$ for $\gamma \geq \gamma_{th}$; note that $\gamma_{th}$ corresponds to the convergence threshold of the iterative decoder [9]. Using (20), we can obtain the waterfall threshold of the turbo code for $L \to \infty$, as follows

$$\gamma_w = \left( \int_{\gamma_{th}}^{\infty} \frac{1}{\gamma^2} d\gamma \right)^{-1} = \gamma_{th}. \qquad (26)$$

Therefore, our approach is in complete agreement with the findings of El Gamal and Hammons [9], i.e., the approximated FER expression is an accurate representation of the actual FER of a turbo code using a long interleaver on a quasi-static fading channel, if the waterfall threshold is set to be equal to the convergence threshold of the iterative decoder.

## VIII. CONCLUSIONS

In this paper, we have considered various transmission schemes, both uncoded and coded, iterative or non-iterative, over quasi-static fading channels and we have demonstrated that a waterfall threshold can be used to characterize them. We have provided an exact interpretation of the waterfall threshold, based on which an accurate approximation of the frame error rate can be obtained. Finally, we have analytically confirmed that our approach is in agreement with the literature, when turbo codes are considered; in particular, we have shown that the waterfall threshold indeed coincides with the convergence threshold of the corresponding iterative decoder, when long interleavers are used.

ACKNOWLEDGMENT

The authors would like to acknowledge the financial support of the Engineering and Physical Sciences Research Council (Grant number: EP/E012108/1).